\documentclass[10pt, a4paper, twocolumn]{article}

\usepackage[hmarginratio=1:1,margin={1.45cm,1.45cm}, top=32mm,columnsep=20pt]{geometry}

\usepackage{subfig}
\usepackage{booktabs}
\usepackage{graphicx}
\usepackage{url}
\usepackage[square,numbers]{natbib}
\usepackage{amsmath}
\usepackage{authblk}

\usepackage{abstract}

\title{Degradation effects of  water immersion on earbud audio quality}

\author[1]{Scott Beveridge}
\author[2,3]{Steffen A. Herff}
\author[1]{Estefan\'{i}a Cano}

\affil[1]{Sonquito, Erlangen, Germany.}

\affil[2]{Digital and Cognitive Musicology Lab (DCML), \'{E}cole Polytechnique F\'{e}d\'{e}rale de Lausanne (EPFL), Lausanne, Switzerland.}

\affil[3]{Music Cognition and Action Research Group (MCA), MARCS Institute for Brain, Behaviour \& Development, Western Sydney University (WSU), Sydney, NSW, Australia.}

\date{}

\begin{document}

\twocolumn[
\begin{@twocolumnfalse}
\maketitle
\begin{abstract}
\noindent Earbuds are subjected to constant use and scenarios that may degrade sound quality. Indeed, a common fate of earbuds is being forgotten in pockets and faced with a laundry cycle (LC). Manufacturers' accounts of the extent to which LCs affect earbud sound quality are vague at best, leaving users to their own devices in assessing the damage caused. This paper offers a systematic, empirical approach to measure the effects of laundering earbuds on sound quality. Three earbud pairs were subjected to LCs spaced 24 hours apart. After each LC, a professional microphone as well as a mid-market smartphone were used to record i) a test tone ii) a frequency sweep and iii) a music signal played through the earbuds. We deployed mixed effects models and found significant degradation in terms of RMS noise loudness, Total Harmonic Distortion (THD), as well as measures of change in the frequency responses of the earbuds. All transducers showed degradation already after the first cycle, and no transducers produced a measurable signal after the sixth LC. The degradation effects were detectable in both, the professional microphone as well as the smartphone recordings. We hope that the present work is a first step in establishing a practical, and ecologically valid method for everyday users to assess the degree of degradation of their personal earbuds.
\end{abstract}
\hspace{0.6cm}
\vspace{0.5cm}
\end{@twocolumnfalse}]

\section{Introduction}

The proliferation of personal media players (PMP) began in the 1970's with Sony's Walkman products.  Advances in digital audio compression technology (like MPEG-1 Audio Layer-3 (mp3) and Advanced Audio Coding (AAC)) combined with cheap solid state storage technology only served to increase PMP popularity in the early 90's. Nowadays the smartphone has superseded the traditional PMP device, with a reported 75\% of consumers using a smartphone to listen to music \cite{musicconsumer18}. 

Although revolutionary with respect to personal music listening, the reduced form factor and portability of PMPs increase the likelihood of accidental damage.  Claim statistics recently published by SquareTrade, a US-based provider of device protection and warranties, report water as the 2nd most likely source of smartphone damage after device dropping \cite{phoneStats}.  Water damage is so frequent that Liquid Contact Indicators (LCIs) have been introduced in many smartphone products, including those from Apple and Samsung.  This is to combat fraudulent warranty claims, as in most cases, liquid immersion or spillage is not covered under the manufacturers' warranty.  

Water damage can come in many forms, and can effect the PMP and device peripherals including the earbuds.  A particular risk for earbuds is being left in pockets of clothing and then exposed to a laundry cycle.  A Google search for ``earbuds water damage'' yielded 63,800 results in the period between the years 2018 to 2019.  The term ``earbuds water damage washing machine'' yields 94,500 in the same time period.  As far as internet searches can be relied upon, accidentally laundering earbuds is a frequent occurrence for many PMP users.

Given that a pair of earbuds have been exposed to water, the primary concerns of earbud owners is to assess the extent of the damage and the effect on audio quality.  In terms of earbud quality, a number of recent smartphone apps promise to assess earbud quality \cite{earphonetestplus, headphonetest}. Whilst conceptually intriguing, these apps often only amount to a selection of test-sounds and still rely on a human listeners to evaluate the signal. Furthermore, such apps may provide insight into the current state of the earbuds but are not able to place them on a degradation trajectory or relate the current state to an interpretable benchmark, e.g., the quality of a new earbud pair. This is understandable, considering that informed estimates of degradation trajectories are earbud model specific and rely on predictive models that are based on real-world degradation data. Here, we collect such data and provide first models of degradation trajectories for the Apple Earpod (model number: A1472).  An Apple product was chosen as it currently represents the most popular brand for consumer earbuds \cite{audioAnalytic19}.  

This paper is focused on assessing the damage caused by water, specifically the process of laundering on earbuds.  We measure this degradation with a pragmatic, ecologically-valid test regime using objective, quantitative measures based on frequency response, harmonic distortion, and noise measures.  Our objective is to examine the degradation trajectory of the earbuds.  In this way, when and if earbuds are damaged, an assessment of degradation can be made and the damaged estimated without relying on listeners expertise.  Of course, full submersion and exposure to the laundry cycle is the extreme case.  

Importantly, we hope the present study will eventually benefit everyday PMP users. Traditionally -and for good reason- audio quality or `health indices' are assessed using professional microphones in order to obtain high quality signals. However, such equipment is often unavailable to everyday users. In order to address this, we record the signals with both, a professional microphone and a simple smartphone. This approach allows us to explore the possibility of using everyday low-quality recordings to obtain earbud health indices. Ultimately, this may lead to a data-driven app, that assesses the health of the earbuds connected to the very same device, simply by placing them in front of the microphone.

\section{Methods}

In this paper, we approach sound quality measurement in a very pragmatic manner by proposing a  simple measurement procedure with equipment readily available to the non-professional.  

\subsection{Test signals}  \label{subsec:testsignals}
We use both synthetic and natural test signals in the measurement procedure as follows: (1) Linear frequency sweep from 20~Hz to 20~kHz, (2) Test tone at 1~kHz, and (3) a pop music track taken from  the Free Music Archive dataset \cite{Defferrard17}.

All test signals are 15 sec long following the ITU-R BS.1387-1 recommendation that states signal duration of natural signals (music) should range between 10 to 20 seconds \cite{ITU-R}.  Although synthetic (sweep/tone) signals can conceivably be much shorter, we chose to match the length of the natural signals.  All test signals were created at a level of -14~dB FS with Audacity audio editor software captured at 32 bit wav format sampled at 44.1 kHz \cite{audacity}.  

\subsection{Measurement equipment} 
The test configuration comprised the following equipment: 
\begin{itemize}
    \item External USB audio interface (Scarlett 2i4, Focusrite, UK).
    \item Condenser measurement microphone (ECM8000, Behringer, Germany) 
    \item Smartphone (Galaxy A5 (2017), Samsung, South Korea)  
    \item Earbuds (x3) (EarPod Headphone Plug (2016) Model number A1472, Apple, USA). 
\end{itemize}

\begin{figure}[t]
\begin{center}
\includegraphics[scale = 1]{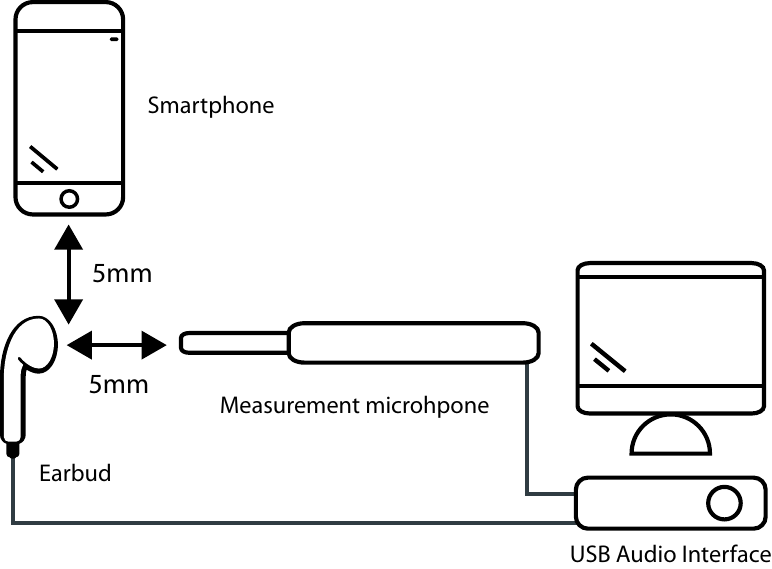}
\caption{Schematic of measurement setup.}
\label{fig:schematicofequipemt}
\end{center}
\end{figure}

\begin{figure}[t]
\begin{center}
\includegraphics[scale = 1.2]{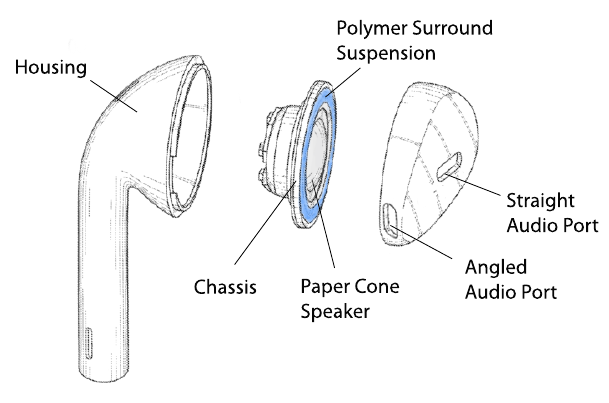}
\caption{The main components of the Apple  Earpod under measurement (adapted  from  \cite{earbuduspat}).  This arrangement is an example of a typical moving coil design.}
\label{fig:explodedearbuds}
\end{center}
\end{figure}

Figure \ref{fig:schematicofequipemt} shows the measurement setup used in this study. The test signals were delivered to the earbuds using Audacity audio editor software via the headphone output of the USB audio interface.  Measurements of the output of the earbuds were made with both the measurement microphone and the smartphone.  The measurement microphone was placed at a fixed distance of 5~mm (see Fig. \ref{fig:schematicofequipemt}) on axis with the angled audio port of the earbud (see Fig. \ref{fig:explodedearbuds}).  The smartphone was placed perpendicular to the earbud angled audio port, also at a distance of 5~mm.  Recordings of the measurement microphone were made with Audacity.  Recordings of the smartphone were captured with the Easy Voice Recorder Android application \cite{easyvoicerecording}.  All recordings were captured in 32 bit wav format sampled at 44.1~kHz. The measurements described in this section are part of a larger body of work that will be reported elsewhere. 

The earpods under test follow a typical moving coil transducer design (see Fig. \ref{fig:explodedearbuds}).  The moving coil transducer converts variations in electrical current into changes in sound pressure.  These sound pressure variations eventually reach the eardrum and can be processed by the brain.  The component responsible for moving the air and creating pressure changes is named the diaphragm.  The diaphragm is the most important, and potentially most delicate part of the moving coil transducer, as it must be both light and strong in order to deliver a full range of frequencies.  The earbuds under test in this paper have a diaphragm of composite design with a paper cone and a polymer surround \cite{ifixit12}.  The polymer surround attaches the paper cone to the chassis.  The surround supports and protects the paper cone while allowing free movement. In the earbuds under test the paper cone faces the straight audio port and is directed to concha cavum on the pinna (the outer ear).  There is an additional angled audio port that projects sound directly into the ear canal (see Fig \ref{fig:explodedearbuds}). The moving coil transducer is  simple electromechanical device susceptible to mechanical wear-and-tear. 

\subsection{Acoustic measurement environment} 
Our aim is to be as pragmatic as possible in our testing regime.  We want to develop a method that tests earbud degradation in an ecologically valid manner.  Measurements were made in a 3~m x 4.5~m room in a residential property.  Ambient sound pressure levels varied between 38.3 and 46.1~dB SPL.  In an attempt to minimize room reflections, we placed our microphones at close proximity to the sound source. We also placed both microphones and the sound source on microphone stands at a height of 1.5~m.  

\subsection{Measurement procedure}
\begin{figure}[!b]
\begin{center}
\includegraphics[scale = 1]{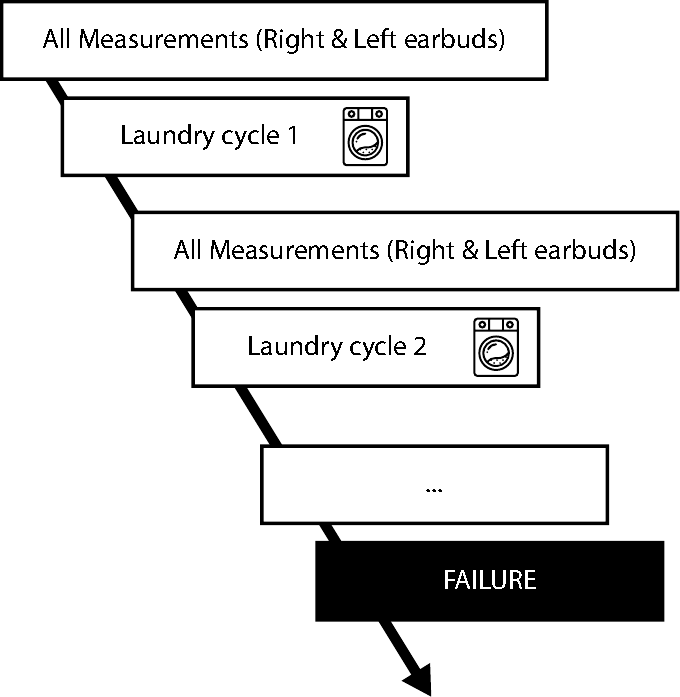}
\caption{Block diagram of experimental procedure}
\label{fig:experimentalProcedureBlock}
\end{center}
\end{figure}

The complete measurement procedure is depicted in Figure \ref{fig:experimentalProcedureBlock}. Three pairs of Apple earbuds (Alpha, Beta, and Gamma) were measured in this study. In each measurement, all the test signals are played through  the earbuds (right and  left), and recorded with both the microphone and smartphone. Each recording is repeated a total of three times (run).
As a degradation stage, the earbuds were then subjected to a laundry cycle. The laundry cycle involves placing each pair of earbuds in a different pocket of a pair of trousers and washing the trousers in a washing machine (Samsung WA85).  The washing machine program was a quick 36 minute normal wash cycle at 28$^{\circ}$C.  The trousers were placed with a number of other garments so that they collectively constituted a half load.  Washing was performed with 40~ml of liquid detergent.  This simulated an every day accidental event of washing the earbuds.  This washing process is referred to as the \textit{Laundry Cycle} (LC).  
Following each LC, the earbuds were left to dry in a room of ambient temperature of 30$^{\circ}$C for a period of 12 hours \cite{Shureprodcutsupport11}.  After this drying period, the earbuds were then measured again and a new LC cycle was performed. The measurement/degradation process was repeated until earbud failure.  

\subsection{Audio quality measures} 
\label{subsec:soundqualitymeasures}
To evaluate the degree of signal and earbud degradation, the following set of quality metrics were calculated after each cycle. 

\subsubsection{Total Harmonic Distortion (THD)}
THD is a measure of the harmonic distortion introduced by a given audio device.  We use a 1~kHz test tone to calculate THD, and consider the first five harmonics in the calculations. The assumption behind THD is that any energy observed in the output signal in harmonic frequencies beyond 1~kHz is the result of distortion introduced by the system. The THD measure is given in dBc (decibels relative to the carrier): a measure of $THD=0~dBc$ occurs when the fundamental frequency and the introduced distortion have the same energy. Negative THD values imply that the energy of the distortion components is less than that of the fundamental. 
For the THD calculations, we use the available Matlab implementation \cite{mathworksthd}.

\subsubsection{PEAQ} 
PEAQ is an algorithm to measure the perceptual quality of audio signals that calculates the difference between basilar membrane representations of the reference and test signals \cite{PEAQ}. Here, we use the Root Mean Square value of the average noise loudness (RMS) as a measure of  difference between reference and test signal. As reference signal, we use the recordings obtained for each earbud pair before any laundry cycle was performed. The free Matlab implementation of PEAQ provided by McGill University was used for the calculations \cite{mmsp}. 

\begin{figure*}[!htb]
\centerline{\includegraphics[scale=0.95]{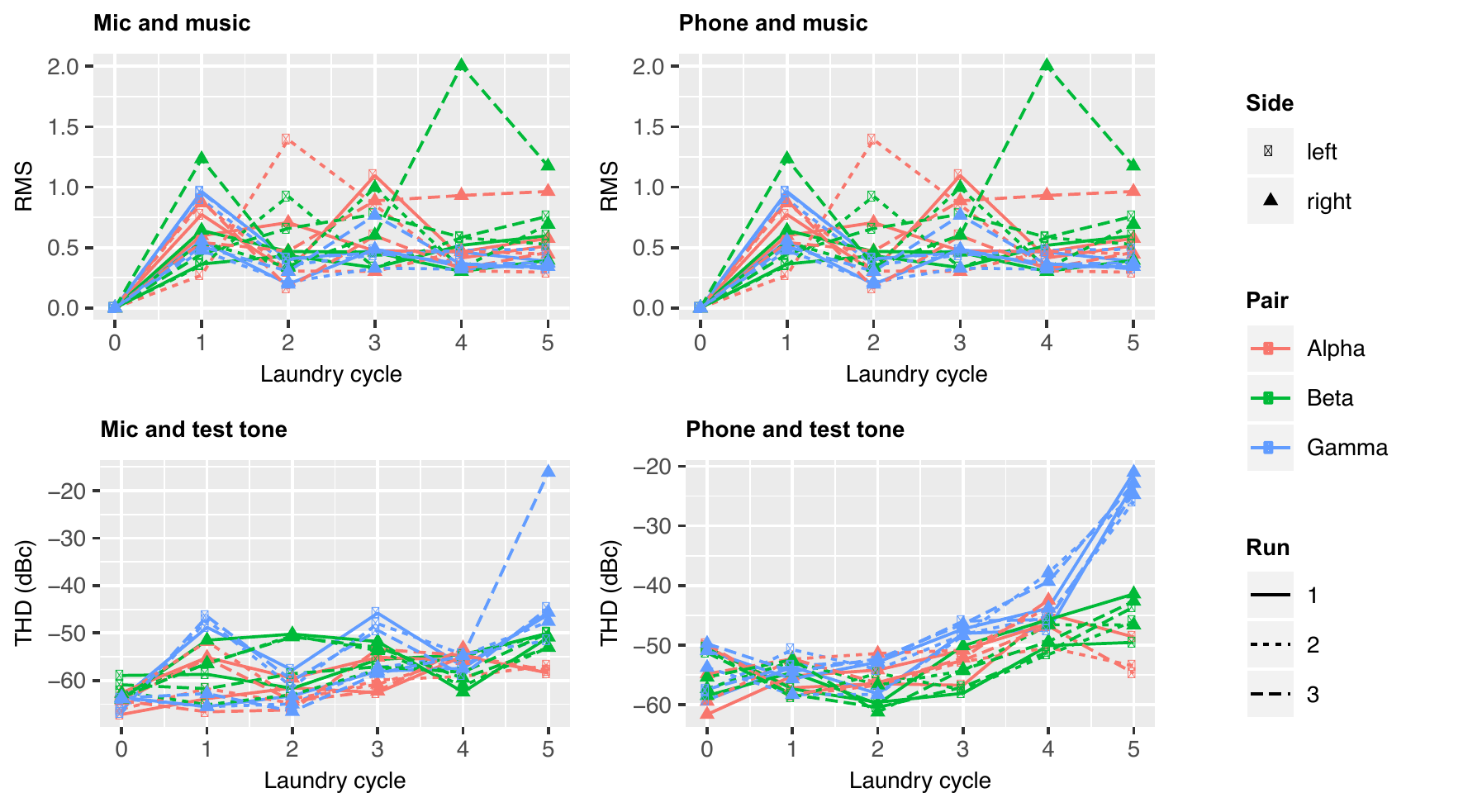}}
\caption{Overview of the THD and RMS values obtained for the test data. The top two panels show the RMS of the average noise loudness obtained for the music signal. The two bottom panels show the corresponding THD values. Both measures are presented independently for the microphone and the phone for all earbud pairs, sides, and runs.}
\label{fig:RMSTHD}
\end{figure*}

\subsubsection{Frequency response}
The frequency response provides a quantitative measure of the spectrum of the earbud in response to an input signal. In this work, we use the linear frequency sweep signal from 20~Hz to 20~kHz to calculate the magnitude of the output as a function of frequency. We use the original frequency response of the  earbuds (as measured with our experimental setup), as a reference to calculate degradation. Any changes observed in the earbuds frequency response after each LC is assumed to be a consequence of earbud decline. 

\section{Results}

\subsection{THD and RMS}
In Figure \ref{fig:RMSTHD}, the THD  values obtained with the test tone, and the RMS values of the average noise loudness obtained with the music signal are provided. The degradation effect of the LCs is clear, with the RMS of the noise loudness increasing with cycle number, as well as the harmonic distortion. Measures for 5 LCs are provided as no signal could be captured with any of the earbuds after LC6. 

\begin{figure}[!hbt]
\centerline{\includegraphics[scale = 0.9]{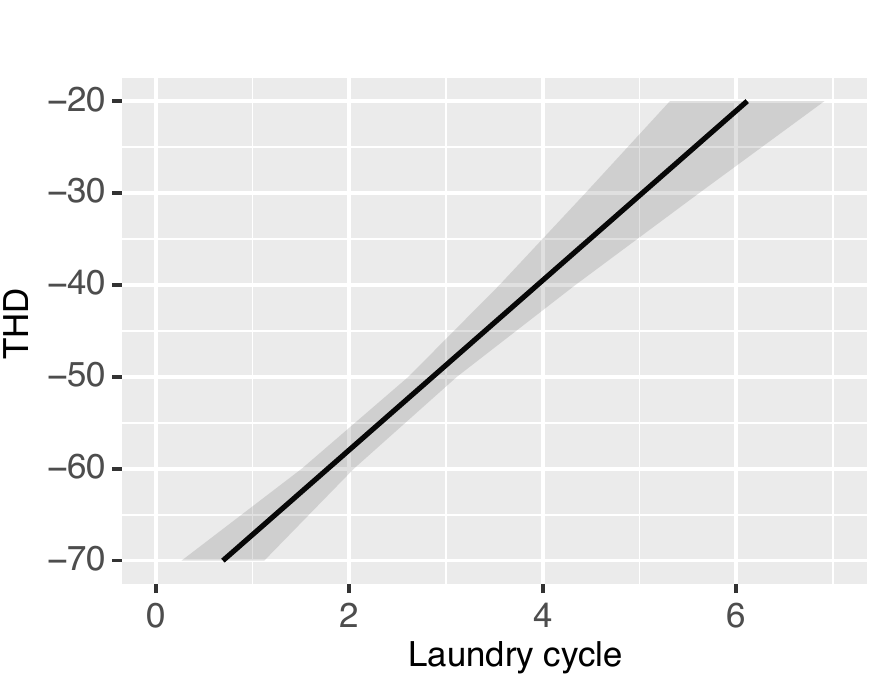}}
\caption{Effect of laundry cycle on THD. The model shows a clear increase in THD with increasing laundry cycle. The model predicts cycle, but for convenience, LC is displayed on the x-axis rather than the y-axis.  The grey band represents a 95\% CI.}
\label{fig:THD}
\end{figure}

To formalize the relationship between LC and degradation, two separate mixed effects models were deployed to predict \textit{laundry cycle} (how many laundry cycles have past) using either a fixed factor for \textit{RMS}, or a fixed factor for \textit{THD}.  Furthermore, the models were provided with a random factor for \textit{Pair:Side} that effectively provides an intercept for each transducer. We report conservative Kenward-Roger adjusted \textit{p}-values \cite{KenwardM1997aaa}. All models were implemented in R using the lme4-package \cite{lme42014}. \textit{RMS} significantly predicted \textit{laundry cycle} (\textit{Est} = 1.9910, \textit{SE} = 0.3341, \textit{p} $<$ .0001), and so did \textit{THD} (\textit{Est} = 0.09073, \textit{SE} = 0.01072, \textit{p} $<$ .0001). Figure \ref{fig:THD} shows that the model captures an increase in THD with increasing cycle number.

\subsection{Frequency responses}
\begin{figure}[!ht]%
\centering
\subfloat[][]{{\includegraphics[scale = 0.9]{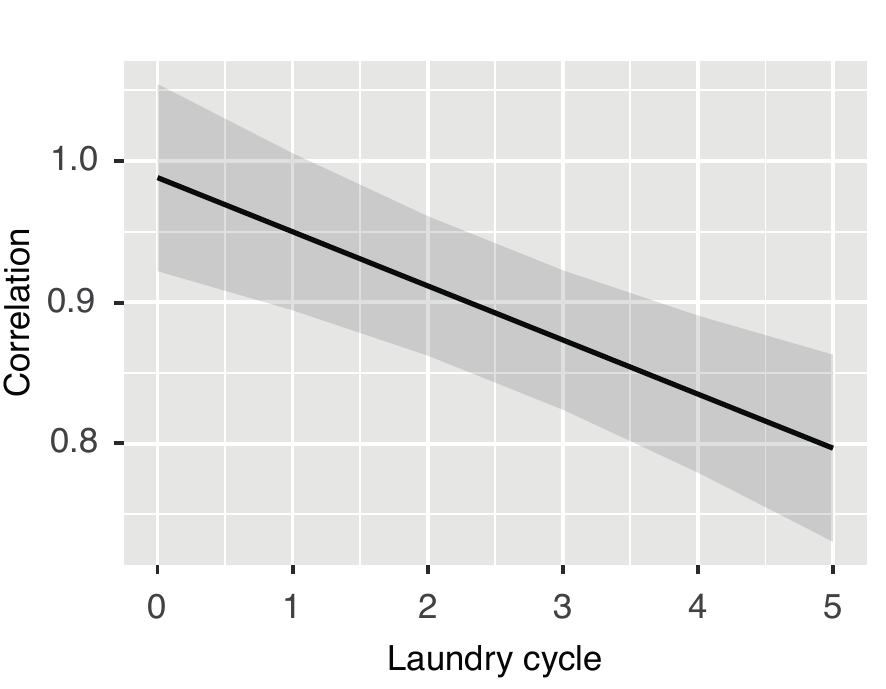}\label{fig:LCvCORR} }}%
\qquad
\subfloat[][]{{ \includegraphics[scale = 0.9]{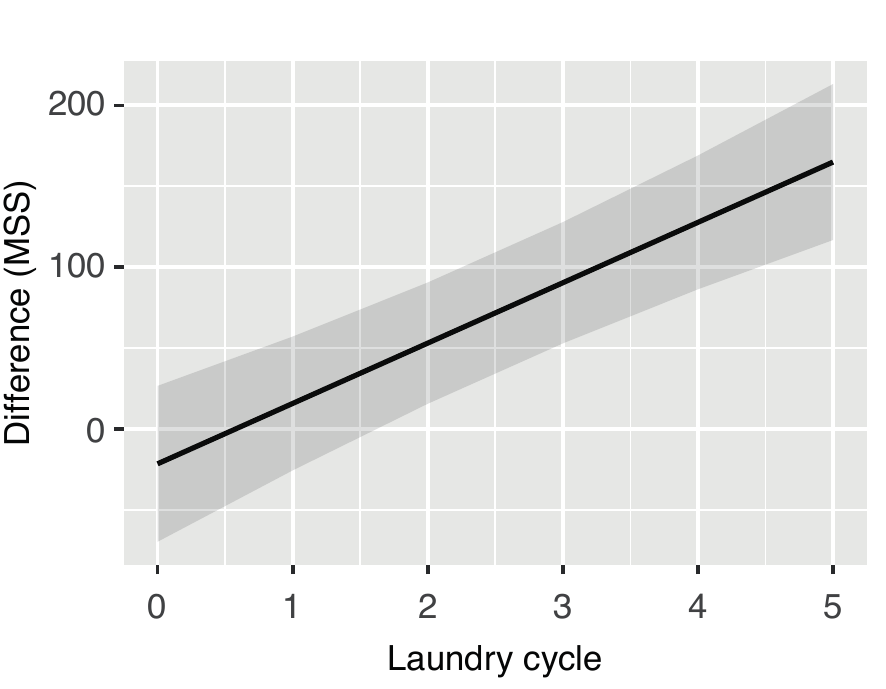}\label{fig:LCvMSS}}}%
\protect\caption{Effect of laundry cycle on the frequency response measures of \protect\subref{fig:LCvCORR} Correlation and \protect\subref{fig:LCvMSS} Difference (MSS). With increasing cycle number, the model shows a clear decrease in the correlation between the frequency responses and a clear increase in the mean difference in magnitudes between the frequency responses. The grey bands represent a 95\% CI.}%
\label{fig:FRMaineffects}%
\end{figure}

The frequency responses of all the earbud pairs, sides, and LCs were calculated using the frequency sweep test signal.  Additionally, to analyze changes in the frequency responses after each LC, two measurements were performed; a correlation measure indicative of changes in the shape of the frequency response, and a difference measure indicative of changes in its magnitude.
First, we calculate the \textit{Correlation} of  the frequency response of each earbud pair, side, and LC with the frequency response obtained from the respective earbud pair and side prior to the first laundry cycle. Second,  we obtain the \textit{Difference} measure by subtracting the frequency response of each earbud pair, side, and LC from the frequency response vector obtained from the respective earbud pair and side, and calculate the mean sum of squares (MSS). To formalize the relationship of these two measures with LC, we deployed two separate mixed effects models predicting the \textit{Correlation} and \textit{Difference MSS} measures using \textit{laundry cycle} and \textit{source} (mic vs smartphone) as fixed effects. Similar to above, the models were provided with a random effect for \textit{side:pair}. As can be seen in Figure \ref{fig:FRMaineffects}, both \textit{Correlation} (\textit{Est} = -0.038283, \textit{SE} = 0.009193 , \textit{p} $<$ .0001) as well as \textit{Difference} (\textit{Est} = 37.295, \textit{SE} = 6.304 , \textit{p} $<$ .0001) were significantly predicted by \textit{laundry cycle}. The recording \textit{source} however, did not significantly predict either (all \textit{p} $>$ .425).

To  examine the potential interaction between \textit{Correlation} and \textit{Difference}, we build an additional mixed effects model predicting \textit{laundry cycle} with the interaction term \textit{Correlation*Difference}. The interaction term significantly (\textit{Est} = 0.020804, \textit{SE} = 0.003989, \textit{p} $<$ .0001) predicted \textit{laundry cycle}. Closer inspection of the marginal effects reveal, that \textit{Difference} only contain predictive information for \textit{laundry cycle}, as long as the \textit{Correlation} is high. In other words, when the frequency response starts to lose its original shape, then the overall difference in magnitude becomes a less useful predictor for degradation. However, as long as the shape is maintained, the overall difference in magnitude is a useful predictor for degradation.

\begin{figure*}[htbp]
\begin{centering}
\subfloat[][]{\label{fig:GammaLeft}\includegraphics[scale = 0.45,trim=1cm 0 0cm 0, clip = true]{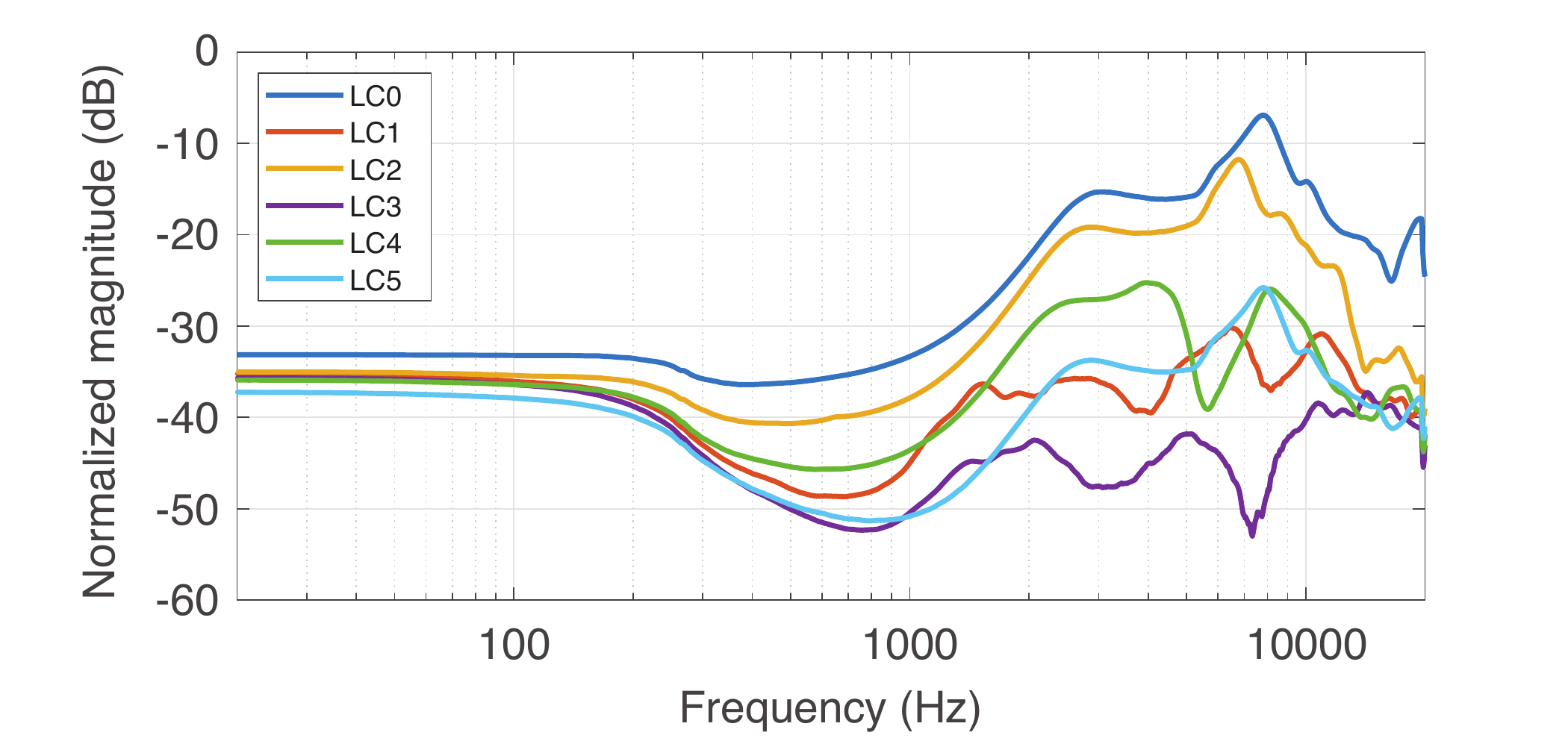}}\subfloat[][]{\label{fig:GammaRight}\includegraphics[scale = 0.45,trim=0cm 0 1cm 0, clip = true]{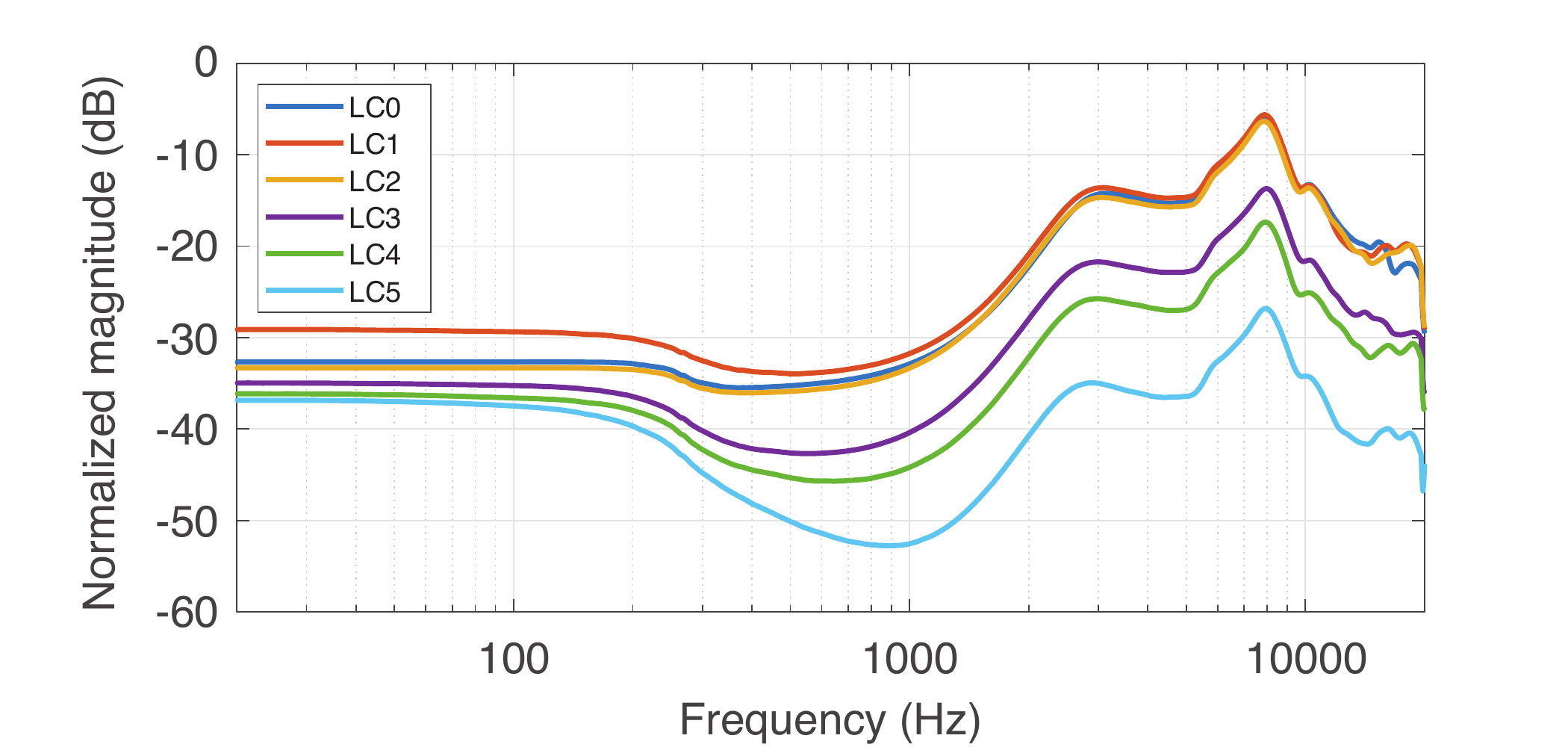}}\caption{\protect\subref{fig:GammaLeft}  Degradation of frequency response of earbud pair \textit{Gamma} left side transducer recorded with the microphone \protect\subref{fig:GammaRight} Degradation of frequency response of earbud pair \textit{Gamma} right side transducer measured with the microphone.  Both transducers show clear  degradation of audio quality as laundry cycles (LC) increase.  In \protect\subref{fig:GammaLeft} as LCs increase frequency specific degradation occurs. in \protect\subref{fig:GammaRight} a more uniform degradation is observed.}
\end{centering}
\end{figure*}

\section{Discussion}

The damage caused by laundry cycles to earbuds has been subject of much debate. Official statements range from `no to minimal destructive effect' \cite{Shureprodcutsupport11} to an immediate voiding of the warranty \cite{applewarranty}. Here, we measure this degradation empirically. We observed that LC cause a substantial damage to earbuds, specifically to three Apple EarPods Headphone (Model number: A1472). The degradation was significant and visible in RMS and THD, as well as in the frequency responses. After six LCs, every single transducer ceased to produce a signal of measurable quantity within the present testing regime. 
In the following, we will discuss the present findings in detail. 
 
\subsection{THD and RMS}

Mixed effects models  showed a significant increase in THD with increasing LCs. These results confirm our initial expectations of progressive degradation with LC, and further allow us to  characterize the extent to which the listener experience can suffer with earbud decline. In the case of THD, degradation can result in changes in the colouring of the sound caused by the introduced distortion \cite{Dobrowohl2019aaa}. The models also showed a significant increase in the RMS of the average noise loudness with increasing LCs. A word of caution is pertinent with respect to the use of PEAQ measures such as the RMS in this context. PEAQ measures are designed to measure perceptual audio quality when a clean, high-quality signal is available as a reference. In our case, we use the initial recordings before any LC, as reference signals for the calculations; however, there is a significant magnitude drop after every LC (as seen in Fig. \ref{fig:GammaLeft} and \ref{fig:GammaRight}), which can have a clear and confounding effect on the RMS of the average noise loudness calculation. 

\subsection{Frequency responses} 

In this work, we use the frequency responses as a measure of degradation and not necessarily as a true representation of the earbud responses (which require a much more rigorous recording setup to be accurately made). With this in mind, we compared the frequency responses after each LC with respect to that of the unwashed earbuds. Specifically, we focused on two measurements, a correlation and a difference measure.

Both the correlation and  difference measures significantly change as the LCs increase. In early laundry cycles, the shape of the frequency response is well preserved in the entire frequency range, in comparison to that of  LC0 (original). A consistent drop in the magnitude of the recorded signal is also evident after each LC. As LC increases, frequency specific degradation become more prevalent. This can for example be seen around 1000 Hz in Figure \ref{fig:GammaRight}, where later cycles affect some frequencies stronger than others. At this point, the overall  magnitude difference between LCs may not be as indicative of the damage, and the degradation is better captured in the correlation parameter.

Figures \ref{fig:GammaLeft} and \ref{fig:GammaRight} exemplify these findings in the frequency responses obtained  for the left and right transducer of the Gamma earbud pair, respectively.

\subsection{Mic vs. Phone}
The present study aimed to take a first step towards a pragmatic, and ecologically valid methodology to assess earbud health. We chose LC as a means to cause degradation, as it is a commonly occurring scenario for everyday PMP users. However, to make a possible future assessment tool of earbud health appealing and generalizable to non-professional users, a proof-of-concept is required to show that professional and expensive recording equipment is not essential. Here, we recorded all signals simultaneously with a professional microphone, as well as a mid-market smartphone. The models deployed were provided with an additional parameter to capture the source of the recording (mic vs phone). The models showed no significant differences between the microphone and the phone recordings when it came to predicting degradation or laundry cycle (LC) number. Both the professional microphone as well as the phone were perfectly capable of capturing the degradation in RMS, THD, and the frequency response measures. We take this as an encouraging finding that modern day smartphones are sufficient to calculate reliable health indices for earbuds. 

\subsection{Causes of degradation}
During degradation cycles earbuds were placed not directly in the washing machine drum, but instead in the in the pockets of a pair of trousers.  At the end of certain cycles earbuds were found outside of the pockets and in the washing machine drum.  In these cases, we observed significant performance degradation in the subsequent measurement phase.  This is especially evident in the frequency response measure of earbud \textit{Gamma} left shown in Figure \ref{fig:GammaLeft}. After LC0 (blue line), this transducer was found outside its assigned pocket and may have sustained additional damage.  This is shown in the reduced magnitude of the frequency response in LC1 (red line).  The frequency response slightly recovered in LC2, but degraded badly on LC3, LC4 and LC5 (purple, green, and cyan lines). 

Given the fragility of the traditional moving coil transducer, escaping from the pocket into the washing machine drum could result in particularly harsh impact to the earbud and cause displacement of the components inside. Both mechanical failure due to impact, or damage caused by increased moisture are possible in this scenario. In general, degradation may come from moisture in the paper cone that makes it slowly degrade over time, detachment of the polymer surround from the cone, detergent saturating the material of the paper cone, etc.  At these small scales, a great number of factors could have an impact on the quality of the earbud.  

In contrast,  the degradation of the frequency response of earbud \textit{Gamma} right (see Figure \ref{fig:GammaRight}) shows  relative uniformity in cycles 0-2 (blue, red, and orange lines) and a slow loss of magnitude in cycles 3-5 (purple, green, and cyan lines). These findings indicate that degradation can significantly differ for each side of the earbud pair.    

\section{Summary}

We empirically tested earbud signal degradation with a pragmatic, and ecologically valid testing regime. We found significant degradation in the signal quality with increasing laundry cycles (LCs), a common fate of everyday earbuds. We observed some transducers recover after quality plunges, however, recovery never achieved initial performance. Anecdotally, placement of the earbuds during the laundry cycle seems to have influenced degradation, with the back pocket of trousers being a particularly hazardous location.

\bibliographystyle{unsrt}

\end{document}